\documentclass[reprint,amsmath,amssymb,aps]{revtex4-1}

\usepackage{graphicx}
\usepackage{dcolumn}
\usepackage{bm}
\usepackage{amssymb}
\usepackage{amsfonts}
\usepackage{amsmath}
\usepackage{float}
\usepackage{siunitx}
\usepackage{multirow}
\usepackage{hyperref}

\begin{document}


\title{Enhanced thermoelectric performance by van Hove singularities in the density of states of type-II nodal-line semimetals}

\author{Nguyen T. Hung$^{1,2}$}
\email{nguyen@flex.phys.tohoku.ac.jp}
\author{Jyesta M. Adhidewata$^{3,4}$}
\author{Ahmad R. T. Nugraha$^{4}$}
\author{Riichiro Saito$^{2}$}

\affiliation{$^1$Frontier Research Institute for Interdisciplinary Sciences, 
Tohoku University, Sendai 980-8578, Japan \\ 
$^2$Department of Physics, Tohoku University, Sendai 980-8578, Japan\\ 
$^3$Department of Physics, Institut Teknologi Bandung, Bandung 40132, 
Indonesia\\
$^4$Research Center for Quantum Physics, National Research and Innovation Agency (BRIN), South Tangerang 15314, Indonesia}

\begin{abstract}
The effects of the unique density of states (DOS) of a topological type-II nodal-line semimetal (NLS) on its thermoelectric (TE) transport properties are investigated through a combination of semi-analytical and first-principles methods with ``spinless Mg$_3$Bi$_2$'' as artificial material. The DOS in such a type-II NLS possesses two van Hove singularities near the energy of the nodal line that leads to a large $S$ value compared to the normal metals.  Combined with the linear band at the nodal line that gives high electrical conductivity $\sigma$, the type-II NLS can exhibit a relatively high TE power factor ($\mathrm{PF}=S^2\sigma$) at the nodal line.  In particular, we find $\text{PF} \sim 60$ \si{\micro}W/cmK$^2$ at 300~K for the n-type Mg$_3$Bi$_2$ by considering the electron-phonon scattering, in which the relaxation time $\tau$ of carriers can be expressed as $\tau\propto\text{DOS}^{-1}$ for the type-II NLS.  Furthermore, we optimize parameters for the TE power factor of type-II NLSs in general by adopting the two-band model with the DOS-dependent relaxation time approximation.  Our results suggest the type-II NLSs as a potential class of high-performance TE materials among metals and semimetals, which are traditionally considered not good TE materials compared to semiconductors. 
\end{abstract}


\date{\today}
\maketitle

Thermoelectric (TE) effects were first reported in metals, in which a voltage difference $\Delta V$ occurs as an electron flows from the hot side to the cold side when a temperature difference $\Delta T$ is applied~\cite{rowe2005thermoelectrics}.  The first measure of thermoelectricity is the Seebeck coefficient, $S=-\Delta V/\Delta T$~\cite{goldsmid2010introduction}.  Searching a material with a high $S$ is thus necessary to develop high-performance TE generators or cooling devices because $S$ is related to the TE device performance characterized by the figure of merit $ZT=S^2\sigma T/(\kappa_e+\kappa_{ph})$~\cite{goldsmid2010introduction}, where $\sigma$ is the electrical conductivity, $\kappa_e$ is the electronic thermal conductivity, $\kappa_{ph}$ is the lattice thermal conductivity and $T$ is the absolute temperature.  Although metals were studied in the early stages of thermoelectrics discovery, their $S$ is usually much smaller than $k_B/e \approx 87$ \si{\micro}V/K, where $k_B$ is the Boltzmann constant and $e$ is the electron charge.  For example, $|S|$ of copper is $6.5$ \si{\micro}V/K at $300$ K. To obtain a better value of $ZT$, people often use doped semiconductors, in which $S$ can be larger than in the metals, on the order of $\mu/k_B T$, where $\mu$ is the energy difference between the Fermi energy and the top (or bottom) of the conduction (or valence) band for the n-type (or p-type) material. For example, $|S|$ of n-type silicon is $340$ \si{\micro}V/K at $300$ K for doping concentration of $2.2\times 10^{18}$ cm$^{-3}$~\cite{geballe1955seebeck}. However, a large $S$ would lead to a small value of $\sigma$ in the semiconductors~\cite{hung2015diameter}. Therefore, finding materials with both large $S$ and high $\sigma$ is the bottleneck to improve the TE power factor, $\text{PF}=S^2\sigma$. In this regard, many studies have been focusing on the PF enhancement for the semiconductors through, for example, band convergence ~\cite{hung2019thermoelectric,tang2015convergence} and quantum confinement~\cite{hicks1993thermoelectric,hicks1993effect,hung2016quantum,hung2021origin}. However, the PF enhancement in the metals or semimetals remains a long-standing problem.

Recently, there has been a lot of attention on materials with non-trivial band topology such as the nodal line semimetals (NLSs)~\cite{shao2020electronic,rudenko2018excitonic}, in which the conduction and valence bands intersect in the form of a line (called the nodal line)~\cite{fu2019dirac}. Based on the slopes of the bands along the nodal line, we can classify the NLSs into type-I and type-II NLSs. The type-I NLS has bands with oppositely aligned slopes near the nodal line, while the type-II NLS has bands with the same slope near the nodal line. Although the semimetals are not suitable for TE applications due to the opposite contribution of electrons and holes to the Seebeck coefficient~\cite{hasdeo2019optimal}, some recent findings pointed out that NLSs might be promising TE materials, such as ZrSiS~\cite{matusiak2017thermoelectric}, Nb$_3$GeTe$_6$~\cite{wang2020unique} or YbMnSb$_2$~\cite{pan2021thermoelectric}.  A previous theoretical study by Zhang et al. suggested that non-trivial electrons in the semimetal ZrTe$_5$ lead to the anomalous Nernst effect and the Seebeck effect~\cite{zhang2019anomalous}, in which both the Seebeck coefficient and Nernst signal exhibit exotic peaks in the strong-field quantum limit. The anomalous transport behaviors of ZrTe$_5$ observed by Shahi et al.~\cite{shahi2018bipolar}. Xu et al.~\cite{xu2014enhanced} also found that the edge states of the topological semimetal lead to the large and anomalous Seebeck effect with an opposite sign with respect to the Hall effect.

\begin{figure}[t!]
\begin{center}
\includegraphics[width=0.45\textwidth]{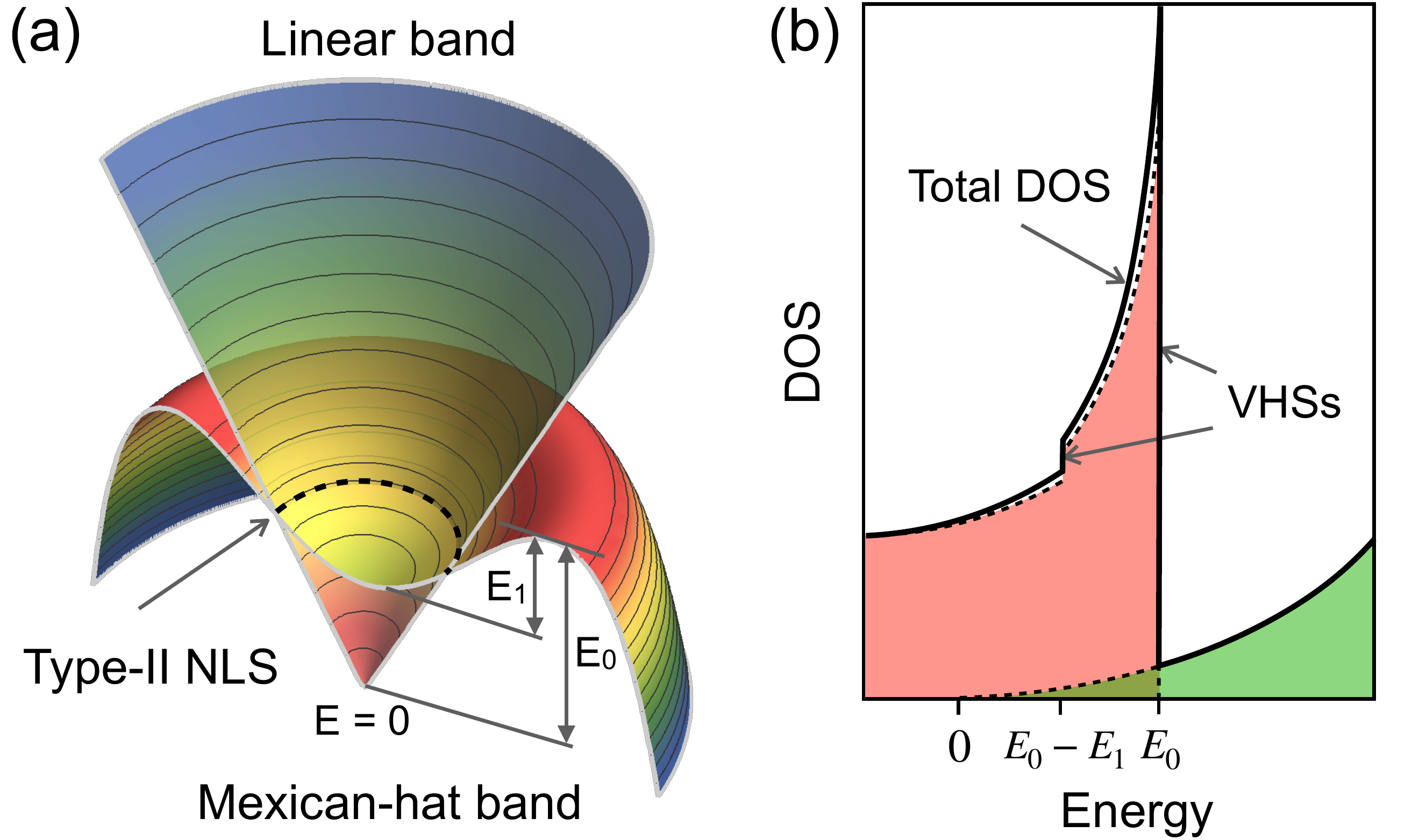}
\caption{Type-II nodal-line semimetal (NLS). (a) Two-band model for type-II NLS with a linear band and a Mexican-hat band. The dashed line denotes the nodal line. (b) Total density of states (solid line) of the linear band (green) and Mexican-hat band (red) has two van Hove singularities (vHSs) at $E_0$ and $E_0-E_1$ [see Eq.~\eqref{eq:dos}].  The dashed lines (with shaded areas) in the DOS plot represent the individual contribution of each band.}
\label{fig:model}
\end{center}
\end{figure}

In this Letter, we investigate the TE response of type-II NLSs using the semi-analytical method by the two-band model (2BM) and the first-principles calculation based on density functional theory (DFT). As an example of the type-II NLSs, we adopt the three-dimensional (3D) Mg$_3$Bi$_2$.  As shown in Fig.~\ref{fig:model}(a), the type-II NLS can be described by a combination of a linear band (LB), $E_c(\bm{k})$, and a Mexican-hat band (MB), $E_v(\bm{k})$, defined by
\begin{equation}
\label{eq:el}
E_c(\bm{k}) = \hbar v_F|\bm{k}|,
\end{equation}
and
\begin{equation}
\label{eq:em}
E_v(\bm{k}) = -\frac{(\hbar^2\bm{k}^2/4m-E_1)^2}{E_1}+E_0,
\end{equation}
respectively, where $\bm{k}$ is the electron wavevector, $v_F$ is the Fermi velocity in the LB, $E_0$ [Fig.~\ref{fig:model}(b)] is a parameter that determines the position of the MB edge with respect to $E=0$, $E_1$ is a parameter which determines the depth of the central valley of the MB measured from the band edge, $m$ is the effective mass of valley in the middle of the MB, and $\hbar$ is the reduced Planck constant. 

From Eqs.~\eqref{eq:el} and~\eqref{eq:em}, the electronic density of states (DOS), $\mathcal{D}_{2b}(E)$, within the two-band model of 3D type-II NLS can be written as (see the Supplemental Material for the derivation):
\begin{widetext}
\begin{equation}
\label{eq:dos}
\mathcal{D}_{2b}(E)=
  \begin{cases}
\frac{2m\sqrt{m E_1}}{\pi^2\hbar^3}\sqrt{\frac{1+\sqrt{(E_0-E)/E_1}}{(E_0-E)/E_1}}       & \quad \text{for } E\leq0,\\
   \frac{E^2}{\pi^2\hbar^3 v_F^3}+
\frac{2m\sqrt{m E_1}}{\pi^2\hbar^3}\sqrt{\frac{1+\sqrt{(E_0-E)/E_1}}{(E_0-E)/E_1}}       & \quad \text{for } 0<E<E_0-E_1,\\
   \frac{E^2}{\pi^2\hbar^3 v_F^3}+
\frac{2m\sqrt{m E_1}}{\pi^2\hbar^3}\left(\sqrt{\frac{1-\sqrt{(E_0-E)/E_1}}{(E_0-E)/E_1}}+\sqrt{\frac{1+\sqrt{(E_0-E)/E_1}}{(E_0-E)/E_1}}\right)   & \quad \text{for }  E_0-E_1 \leq E < E_0,\\
   \frac{E^2}{\pi^2\hbar^3 v_F^3}   & \quad \text{for }  E \geq E_0.\\
  \end{cases}
\end{equation}
\end{widetext}
Equation~\eqref{eq:dos} shows two van Hove singularities (vHSs): one from the divergence of the DOS at $E=E_0$ and the other from the discontinuity at $E=E_0-E_1$, as shown in Fig.~\ref{fig:model}(b). The nodal line exists at the energy between the two vHSs. Therefore, the enhancement of DOS occurs in a small region between the two vHSs for the type-II NLS. The vHSs also enhance the PF of the low-dimensional materials due to the quantum confinement effect~\cite{hicks1993effect,hung2016quantum,hung2021origin}. On the other hand, the LB generally gives a large $\sigma$ as reported in the cases of graphene~\cite{stoller2008graphene} and three-dimensional Dirac Cd$_3$As$_2$~\cite{neupane2014observation}.


\begin{figure*}[ht!]
\begin{center}
\includegraphics[width=0.95\textwidth]{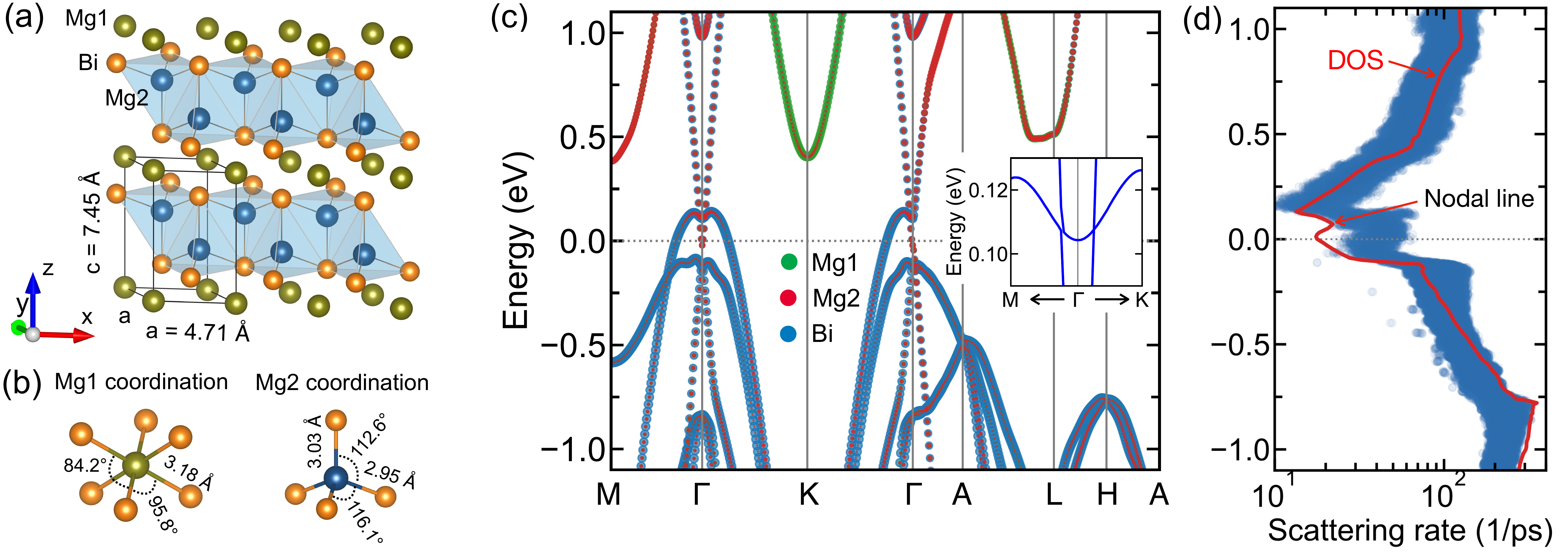}
\caption{Mg$_3$Bi$_2$ as an example of type-II NLSs. (a) Crystal structure of Mg$_3$Bi$_2$. The orange, green, and blue balls indicate the Bi, Mg1, and Mg2 atoms, respectively. (b) Coordination polyhedrons of the non-equivalent Mg1 and Mg2 atoms. (c) Electronic band structure with atomic orbital projections in color. The inset shows a tiny gap and a type-II Dirac point along the $\Gamma$-M and $\Gamma$-K directions, respectively. (d) Electronic density of states (DOS) and scattering rate at 300 K as a function of energy. A local increase in DOS occurs near the nodal line at $\sim 0.1$ eV.}
\label{fig:mg3bi2}
\end{center}
\end{figure*}

In order to show an example of the type-II NLS, we adopt 3D Mg$_3$Bi$_2$ whose electronic structure and phonon-related properties are calculated with the first-principles DFT method.  We use Quantum ESPRESSO~\cite{giannozzi09-espresso} to perform the DFT calculations with the Perdew-Burke-Ernzerhof parameterization~\cite{perdew1996generalized} for the exchange-correlation functional under the generalized gradient approximation.  Ultrasoft pseudopotentials with a kinetic energy cutoff of 60 Ry are also employed in the calculations. We then use the EPW package~\cite{ponce2016epw} to calculate the carrier scattering rate, which is given by the inverse of the relaxation time $\tau_{m\mathbf{k}}^{\mathrm{el-ph}}$ by the electron-phonon interaction as
\begin{equation}
\label{eq:tau}
\begin{split}
\frac{1}{\tau_{m\mathbf{k}}^{\mathrm{el-ph}}}=&\frac{2\pi}{\hbar}\sum_{m'\nu j}\int\frac{d\mathbf{q}}{\Omega_{\text{BZ}}}|\mathcal{M}(\mathbf{k},\mathbf{q})|^2\times\\
&\left[\frac{1+j}{2}-jf_{m'\mathbf{k}+\mathbf{q}}^0+n_{\mathrm{q}\nu}\right]\delta(\Delta E_{{\mathbf{k},\mathbf{q}}}-j\hbar\omega_{\mathbf{q}\nu}),
\end{split}
\end{equation}
where $\mathcal{M}(\mathbf{k},\mathbf{q})$ is the electron-phonon matrix element for the scattering of an electron from the initial state $m\mathbf{k}$ to the final state $m'\mathbf{k}+\mathbf{q}$ by emitting (or absorbing) a $\nu$-th phonon with the wavevector $\mathbf{q}$ and frequency $\omega_{\mathbf{q}\nu}$.  We consider all electron-phonon scattering processes, including acoustic, optical, and polar-optical phonons for $\mathcal{M}(\mathbf{k},\mathbf{q})$.  In Eq.~\eqref{eq:tau}, $f_{m'\mathbf{k}+\mathbf{q}}^0$ and $n_{\mathrm{q}\nu}$ are the Fermi-Dirac and Bose-Einstein distribution functions, respectively.  The index $j$ takes a value of $\pm 1$ for phonon emission ($+$) and absorption ($-$), respectively. The electron energy and phonon dispersion relations in Eq.~\eqref{eq:tau} are initially calculated on $10\times10\times6$ $\mathbf{k}$-point and $5\times5\times3$ $\mathbf{q}$-point grids, respectively. Then, we interpolate the electronic energy and phonon dispersions on the dense meshes of $70\times70\times42$ $\mathbf{k}$-point and $35\times35\times21$ $\mathbf{q}$-point grids for calculating $\tau_{m\mathbf{k}}^{\mathrm{el-ph}}$, respectively. 

In Fig.~\ref{fig:mg3bi2}(a), we show the crystal structure of Mg$_3$Bi$_2$, in which the lattice is centrosymmetric with space group of $P\bar{3}m1$. The calculated lattice constants by geometry optimization are $a=b=4.683$ \AA\ and $c=7.396$ \AA\ and reproduce the experimental ones~\cite{watson1984electronic} ($a=b=4.666$ \AA\ and $c=7.401$ \AA) with the deviation $<0.4$ \%. The Mg$_3$Bi$_2$ structure can be described by the ionic layers of the Mg1 atoms sandwiched by two-dimensional (2D) covalent layers of the Mg2 and Bi atoms. The Mg1 and Mg2 atoms are non-equivalent atoms with different coordination, as shown in Fig.~\ref{fig:mg3bi2}(b). The Mg1 atom is connected to six Bi atoms with six equal bond lengths of $3.18$ \AA, while the Mg2 atom is tetrahedrally coordinated by the four Bi atoms with three shorter bond lengths of $2.95$ \AA\ and one vertical bond length of $3.03$ \AA. Therefore, the Mg1 and Mg2 atoms contribute differently to the electronic band structure.

In Fig.~\ref{fig:mg3bi2}(c), we show the electronic band structure of Mg$_3$Bi$_2$ with atomic orbital projections in color. Here we do not consider spin-orbit interaction. Near the Fermi energy ($E = 0.11$ eV) and along the $\Gamma$-M and $\Gamma$-K directions, two bands characterized by the LB and MB [see Fig.~\ref{fig:model}(a)] cross each other, indicating a semimetallic property in Mg$_3$Bi$_2$. The atomic orbital projections indicate that the LB and MB originate from the Mg2 and Bi atoms of the 2D covalent layer, respectively. By zooming the band overlap region [inset of Fig.~\ref{fig:mg3bi2}(c)], we find that the crossing point is gapless along the $\Gamma$-K direction, and the slopes of the two bands at the crossing point have the same sign, indicating a type-II Dirac point (DP). On the other hand, along the $\Gamma$-M direction, there is a tiny gap ($\sim 1$ meV) between the two bands. Quan et al.~\cite{quan2017single} demonstrated that in a spinless system with both space-inversion and time-reversal symmetries, the crossing points must form a one-dimensional (1D) line (or nodal line) rather than discrete points. In our calculation, the nodal line around the $\Gamma$ point contains the type-II DP along the $\Gamma$-K direction. However, the nodal line does not exactly lie on the $\Gamma$-M-K plane but slightly wiggles around this plane. Therefore, a tiny gap exists along the $\Gamma$-M direction. This result is consistent with previous experimental and theoretical studies~\cite{chang2019realization,zhang2017topological}. It is important to note that the type-II NLS occurs in Mg$_3$Bi$_2$ only for the spinless case. When we take into account the spin-orbit interaction in the band calculation, an indirect gap of 35 meV appears along the $\Gamma$-M direction (see Fig. S4 in the Supplemental Material). Thus, real Mg$_3$Bi$_2$ is no longer the type-II NLS. Since the present study focuses on the conventional TE properties of the type-II NLS, we consider only the spinless Mg$_3$Bi$_2$ case. Our calculated results do not correspond to real Mg$_3$Bi$_2$ with spin-orbit coupling, yet it can give insight into desired enhancement of TE properties in the real system.

In Fig.~\ref{fig:mg3bi2}(d), we show the DOS (red line) and the scattering rate ($1/\tau^{\mathrm{el-ph}}$) (blue region) at 300 K of the spinless Mg$_3$Bi$_2$ as a function of the energy. The enhancement of DOS is found at the energy of nodal line ($\sim 0.1$ eV) due to the vHSs, as discussed by the two-band model. The enhancement of DOS has the window energy about 0.125 eV, which is much larger than $k_BT=0.025$ eV at the room temperature. Therefore, the electrons can exist on the enhancement of DOS even for the room temperature. The scattering rate shows a value of relaxation time $\tau^{\mathrm{el-ph}}\approx 0.03$ ps at the nodal line. The energy dependence of the scattering rate suggests that $\tau^{\mathrm{el-ph}}$ can be expressed by the DOS as $\tau^{\mathrm{D}}=C/\text{DOS}$, where $C = 2.2\times 10^{-13}$ is a constant in units of $\mathrm{sec}\times\text{state}/\text{Ry}/\text{volume}$, which is obtained by the least-squares fitting in BoltzTraP2~\cite{madsen2018boltztrap2}.

Using calculated band structure and the relaxation time, $S$, $\sigma$, and $\kappa_e$ are obtained within the linearized Boltzmann theory as~\cite{hung2018universal}
\begin{equation}
\label{eq:te1}
S=\frac{1}{qT}\frac{\mathcal{L}_1}{\mathcal{L}_0},~~\sigma=q^2\mathcal{L}_0,~~\kappa_e=\frac{1}{T}\left(\mathcal{L}_2-\frac{\mathcal{L}_1^2}{\mathcal{L}_0}\right),
\end{equation}
where the transport integrals $\mathcal{L}_i$ with $i=0,1,2$ are defined by
\begin{equation}
\label{eq:te2}
\mathcal{L}_i=\int{v_x^2\tau(E)\mathcal{D}(E)(E-\mu)^i\left(-\frac{\partial f_0}{\partial E}\right)dE},
\end{equation}
where $v_x$ is the group velocity in the $x$ direction. Eqs.~\eqref{eq:te1} and~\eqref{eq:te2} are calculated by using the BoltzTraP2 package~\cite{madsen2018boltztrap2}. We consider three scattering models for the carrier relaxation-time approximation (RTA): (1) $\tau(E)$ within the electron-phonon coupling approximation (EPA), represented by $\tau(E)=\tau^{\mathrm{el-ph}}$, (2) $\tau(E)$ inversely proportional to the total DOS (DRTA) of all bands, denoted by $\tau(E)=C/\mathcal{D}$, and (3) the constant RTA (CRTA), represented by $\tau(E)=\tau_0$ with $\tau_0$ is set to $3\times 10^{-14}$ s.  For further insight into the relevant scattering mechanisms, we also calculate the TE properties of Mg$_3$Bi$_2$ semi-analytically by using the two-band model with $\tau(E)=C_0/\mathcal{D}_{2b}$, where $\mathcal{D}_{2b}$ is given in Eq.~\eqref{eq:dos}, and $C_0=3.2 \times 10^{-14}$ s/Ry/volume . The detailed derivation and the fitting parameters for the two-band model can be found in the Supplemental Material.

\begin{figure}[ht!]
\begin{center}
\includegraphics[width=0.45\textwidth]{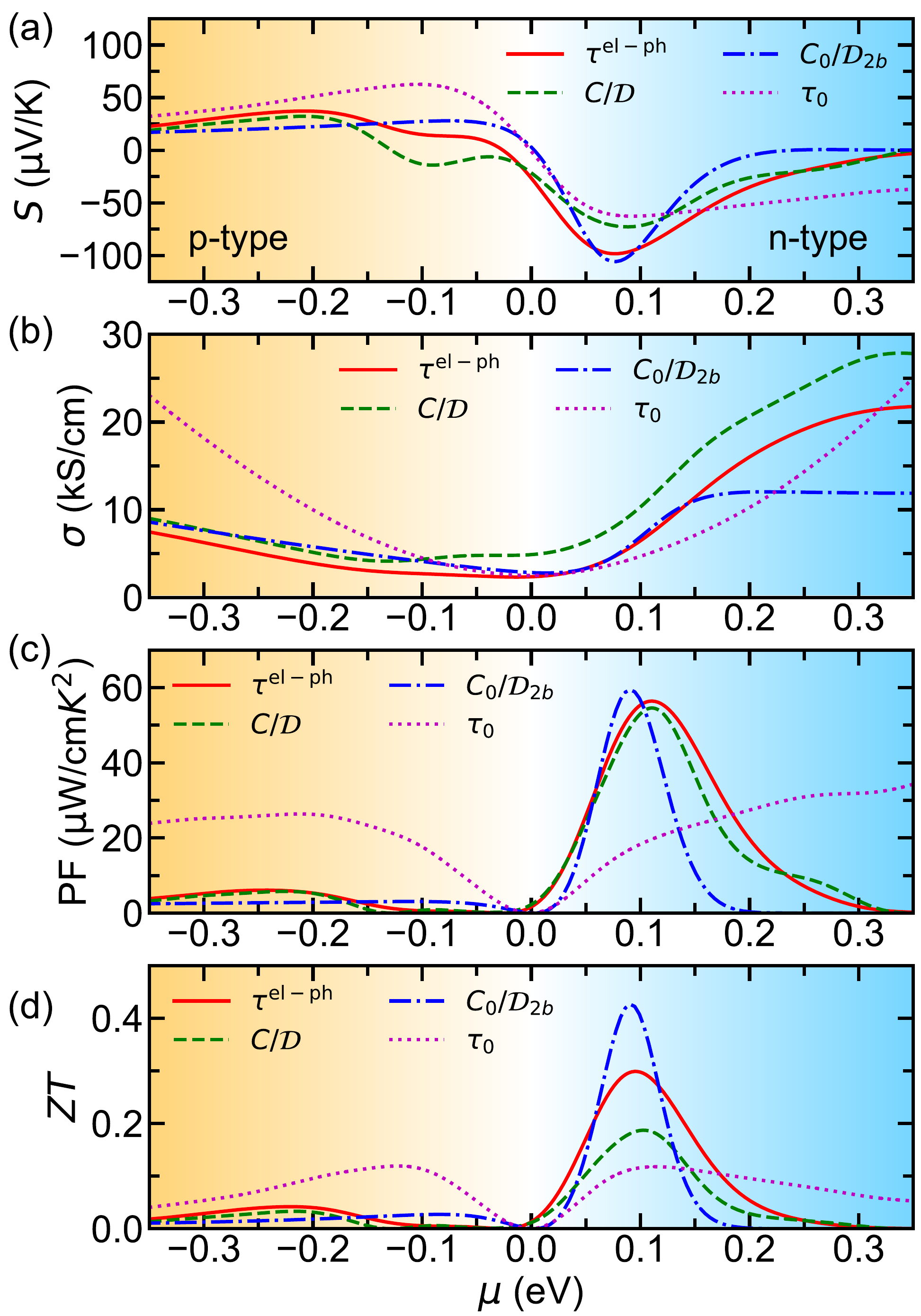}
\caption{(a) Seebeck coefficient $S$, (b) electrical conductivity $\sigma$, (c) thermoelectric power factor PF, and (d) figure of merit $ZT$ of spinless Mg$_3$Bi$_2$ NLS are plotted as a function of the chemical potential $\mu$ at 300 K. Four scattering-models are calculated for the relaxation-time approximation, including the EPA ($\tau(E)=\tau^{\mathrm{el-ph}}$), the DRTA ($C/\mathcal{D}$ and $C_0/\mathcal{D}_{2b}$), and CRTA ($\tau_0$).
Yellow and blue regions denote the p-type and n-type, respectively. The optimized PF and $ZT$ values are found at the nodal line $\mu \sim 0.1$ eV for n-type Mg$_3$Bi$_2$.}
\label{fig:pf}
\end{center}
\end{figure}

In Figs.~\ref{fig:pf}(a)-(d), we show $S$, $\sigma$, PF, and $ZT$ for the $x$ direction of Mg$_3$Bi$_2$ as a function of $\mu$ at $T=300$ K, respectively. In present study, we focus on the in-plane TE properties since Mg$_3$Bi$_2$ has the layer structure, as shown in Fig.~\ref{fig:mg3bi2}(a). It is noted that $S$ and $\sigma$ in the $y$ direction are similar with that in the $x$ direction. As shown in Figs.~\ref{fig:pf}(a) and (b), the DRTA (i.e., $\tau(E)=C/\mathcal{D}$ and $C_0/\mathcal{D}_{2b}$) can reproduce the asymmetry  (with respect to p- and n-type regions) of $|S|$ and $\sigma$ found in the case of the EPA (i.e., $\tau(E)=\tau^{\mathrm{el-ph}}$). The asymmetry of TE properties can be explained by the term of $v^2_x$ in Eq.~\eqref{eq:te2} when the term of $\tau(E)\mathcal{D}(E)$ becomes a constant ($C$ or $C_0$) for the DRTA cases. On the other hand, the CRTA (i.e., $\tau(E)=\tau_0$) shows almost symmetric $S$ and $\sigma$ with respect to p-type and n-type regions.  This result suggests that the CRTA is not suitable to describe the TE properties of the type-II NLS even though it is widely used for TE properties calculation for semiconductor. 

The optimized value of $|S|$ is found to be 100 \si{\micro}V/K at $\mu=0.08$ eV for the n-type, as shown in Fig.~\ref{fig:pf}(a), which is higher than upper limit of the metal ($k_B/e\approx 87$ \si{\micro}V/K). For graphene, the maximum $|S|$ is reported about 60 \si{\micro}V/K for the p-type~\cite{kanahashi2019giant}. The large $|S|$ of Mg$_3$Bi$_2$ at near the nodal line can be explained by the enhancement of DOS, as shown in Eq.~\eqref{eq:dos}. On the other hand, we have $\sigma\propto \mathcal{L}_0\propto v_x$ from Eqs.~\eqref{eq:te1} and~\eqref{eq:te2}) within the EPA and DRTA. Therefore, $\sigma$ for the n-type is larger than that for the p-type because of the contribution of the LB, as shown in Fig.~\ref{fig:pf}(b). 

Based on $S$ and $\sigma$, the optimized PF $\sim 55$ \si{\micro}W/cmK$^2$ is found at $\mu\sim 0.1$ eV for the n-type Mg$_3$Bi$_2$, which is much larger than the p-type case. This optimized PF is larger than the experiment value for the n-type Mg$_{3.2}$Bi$_{2-x}$Sb$_{x}$ at room temperature ($\sim 25-28$ \si{\micro}W/cmK$^2$~\cite{mao2019high,liang2021high}). Zhang et al.~\cite{zhang2019fermi} also reported the PF $\sim 14$ \si{\micro}W/cmK$^2$ for the n-type Mg$_3$Bi$_2$ with spin-orbit coupling within the CRTA. We note that due to the spin-orbit interaction and the doping effect, Mg$_3$Bi$_2$ shows a transition from the NLS to metal or semiconductor, which might lead to reducing the PF compared with the spinless case.

In Fig.~\ref{fig:pf}(d), we show the $ZT=\text{PF}/(\kappa_e+\kappa_{ph})$ value as a function of $\mu$ at $T=300$ K. Here, we adopt $\kappa_{ph}=1$ W/mK from the experimental data~\cite{mao2019high} of Mg$_3$Bi$_2$, and the calculated $\kappa_e$ is around 5 W/mK, as shown in Fig. S5 in the Supplemental Material. The optimized $ZT \sim 0.3$ is found at $\mu\sim 0.1$ eV for the EPA case. Although $ZT$ of the NLS is smaller than the TE semiconductor, the high PF of the NLS is important for the specific TE application. In particular, when the heat source is unlimited, such as solar heat, the output power density, $Q\propto\text{PF}$, is an essential parameter to evaluate the performace and TE generation cost (\$ per W)~\cite{yee2013per,liu2015n-type}.

Using the two-band model, we show the optimized value of the PF as a function of $E_1$ and $E_0$ (see Fig. S3 in the Supplemental Material). We suggest that the best type-II NLS for the TE should have a small $E_0-E_1$ on the order of a few meV, but a condition of $E_1 < E_0 \neq 0$ should be satisfied to keep the material as the type-II NLS. We also calculate the TE properties of Mg$_3$Bi$_2$ with different temperatures (200K, 300K, 400K, and 500K, see Fig. S6 in the Supplemental Material), which shows that the PF increases with decreasing the temperature. This suggests Mg$_3$Bi$_2$ as a potential n-type TE material for low-temperature energy harvesting.

In conclusion, we have shown that the enhancement of DOS of type-II NLS leads to a high Seebeck coefficient at the nodal line region, which can be higher than the upper limit of the metal $k_B/e \approx 87$ \si{\micro}V/K. On the other hand, the linear band of the type-II NLS also contributes to high electrical conductivity. Therefore, we can obtain an optimized power factor with the chemical potential at the nodal line.  The unique DOS of the type-II NLS plays an important role in improving the TE properties of the semimetal.

N.T.H. acknowledges JSPS KAKENHI (Grant No. JP20K15178) and the financial support from the Frontier Research Institute for Interdisciplinary Sciences, Tohoku University.  J.M.A. and A.R.T.N. performed some of the calculations with the help of HPC LIPI facility.  R.S. acknowledges JSPS KAKENHI Grant No JP18H01810.

%

\end{document}